\documentstyle[PASJadd,epsf]{PASJ95}
\begin{document}
\setcounter{page}{1}

\title{Stability of Dynamically Collapsing Gas Sphere}

\author{Tomoyuki {\sc Hanawa} \\
{\it Department of Astrophysics, Nagoya University, Chikusa-ku, Nagoya
464-8602} \\
{\it  E-mail(TH): hanawa@a.phys.nagoya-u.ac.jp} 
and \\
Tomoaki {\sc Matsumoto} \\
{\it Faculty of Liberal Arts, Hosei University,
Fujimi, Chiyoda-ku, Tokyo 102-8160} \\
{\it  E-mail(TH): matsu@i.hosei.ac.jp} 
}

\abst{We discuss stability of dynamically collapsing
gas spheres.   We use a similarity solution for a dynamically
collapsing sphere as the unperturbed state.  In the
similarity solution the gas pressure is approximated by
a polytrope of $ P \, = \, K \rho ^\gamma $. 
We examine three types of perturbations: 
bar ($ \ell \, = \, 2$) mode, spin-up mode, and 
Ori-Piran mode.   
When $ \gamma \, < \, 1.097 $,
it is unstable against bar-mode.  It is unstable 
against spin-up mode for any $ \gamma $.
When $ \gamma \, < \, 0.961 $, the
similarity solution is unstable against Ori-Piran mode.
The unstable mode grows in proportion to 
$ \vert t \, - \, t _0 \vert ^{-\sigma} $ while the central
density increases in proportion to 
$ \rho _c \, \propto \, (t \, - \, t _0) ^{-2} $ 
in the similarity solution.  
The growth rate, $ \sigma $ is obtained numerically
as a function of $ \gamma $ for bar mode and
Ori-Piran mode.
The growth rate of the bar mode is larger for a smaller
$ \gamma $.
 The spin-up mode has the growth
rate of $ \sigma \, = \, 1/3 $ for any $ \gamma $.}
\kword{Gravitation --- Hydrodynamics --- ISM:
clouds --- Stars: formation}

\maketitle

\section{Introduction}

As well as the majority of nearby pre-main sequence stars,
most young stars have their companions (see, e.g., Mathieu
1994).  Since even young protostars have their companions,
fragmentation is more plausible for formation of multiple
star systems than capture and other processes. 

Fragmentation during the star formation has been paid
much attention and a number of numerical simulations 
have been performed for clarifying it.  Nonetheless the
results are not conclusive.  Some early simulations have
only limited spatial resolution and their results on
fragmentation are unreliable (Truelove et al. 1997).  
A gas cloud fragments artificially in numerical simulations 
of low resolution.  If we exclude the artificial fragmentation,
fragmentation proceeds but only slowly in numerical simulations.
Fragmentation competes with the total collapse of a cloud.
A cloud collapses and the density increases before it fragments.
Only when an initial model was very much elongated,
the simulation showed fragmentation of a cloud at a relatively
low density. 
This initial model is, however, not applicable to a roundish
dense core containing a binary.

Fragmentation has been studied not only by numerical simulations
but also by linear stability analyses.  While numerical simulations
are advantageous at handling nonlinearity, linear stability 
analyses are helpful for elucidating underlying physics.  
In this sense these two technologies are complementary.
Most linear stability analyese thus far, however, deal with
the stability of an equilibrium cloud against small perturbation.
The competition with the total collapse could not be taken 
account into them.  Silk, Suto (1988) tried to compute 
the stability of a similarity solution for the collapse of
an isothermal gas cloud
but could not obtain a self-consistent solution.
Only recently Hanawa, Matsumoto (1999) 
succeeded in analyzing the stability 
of a collapsing isothermal cloud against non-spherical
perturbations.   They found that 
a collapsing isothermal cloud is unstable against bar mode
($ \ell \, = \, 2 $) perturbation.  The bar mode grows
in proportion to $ \rho _{\rm c} {}^{0.354} $ where $ \rho _{\rm c} $
denotes the central density.  When it grows, a collapsing
gas cloud may change its shape from sphere to filament.
This result is consistent with recent three-dimensional
simulations by Truelove et al. (1998) and Matsumoto, Hanawa
(1999) in which a filament forms in a collapsing cloud.
The former showed that the filament fragments to form a
binary.  Thus the bar mode is interesting for understanding
fragmentation during the collapse. 

In this paper we extend the linear stability analyses of
Hanawa, Matsumoto (1999) for collapsing non-isothermal spheres.
For simplicity we employ the polytropic relation, 
$ P \, = \, K \rho ^\gamma $, for the model cloud.  
The polytrope model can describe  temperature change during the
collapse in the most simple form.  The polytrope model of
$ \gamma \, \simeq \, 4/3 $ can be applied also to a collapsing
iron core resulting Type II supernova.
In section 2 we present the similarity solution for
gravitationally collapsing sphere of a polytropic gas cloud.
This section is essentially the review of Yahil (1983) and
Suto, Silk (1991).  We summarize their results to investigate
the stability.  In section 3 we examine the stability of
the similarity solution against bar mode.  The growth rate
of the perturbation is larger when $ \gamma $ is smaller.
The bar mode is degenerate and its growth rate is independent
of the azimuthal wavenumber, $ m $.  The shape of the core
suffering the bar mode depends on $ m $ and the sign
of the perturbation.  We show it visually.
In section 4 we show that a spherically collapsing gas cloud is unstable
against spin-up.  This result is an extention of the
stability analysis of Hanawa, Nakayama (1997).
In section 6 we show that the similarity solution is unstable
against a spherical perturbation when $ \gamma \, < \, 0.961 $.
The analysis shown in section 5 is based on Ori, Piran (1988).
A short summary is given in section 6. 

\section{Similarity  Solution}

For simplicity we consider gas of which equation of state
is expressed by polytrope,
\begin{equation}
P \; = \; K \, \rho ^\gamma \; , \label{polytrope}
\end{equation}
where $ P $ and $ \rho $ denote the pressure and density,
respectively. 
The hydrodynamical equations are then expressed as
\begin{equation}
\frac{\partial \rho}{\partial t} \; + \;
\nabla \cdot ( \rho \mbox{\boldmath$v$} ) \; = \; 0 \;
\end{equation}
and 
\begin{equation}
\frac{\partial}{\partial t} ( \rho \mbox{\boldmath$v$} )
\; + \; \nabla P \; + \; \nabla \cdot
(\rho \mbox{\boldmath$v$} \otimes \mbox{\boldmath$v$}) \;
+ \; \rho \nabla \Phi \; = \; 0 \; ,
\end{equation}
where $ \mbox{\boldmath$v$} $ and $ \Phi $ denote the
velocity and gravitational potential, respectively.
The gravitational potential is related with the density
distribution by the Poisson equation,
\begin{equation}
\Delta \Phi \; = \; 4 \pi G \rho \; ,
\label{poisson}
\end{equation}
where $ G $ denotes the gravitational constant.

For later convenience, we introduce the zooming coordinates
of Bouquet et al. (1985) to solve equations (\ref{polytrope})
through (\ref{poisson}). 
The zooming coordinates,
$ (\mbox{\boldmath$\xi$}, \, \tau) $, are related with the
ordinary coordinates, $ (\mbox{\boldmath$r$}, \, t) $, by
\begin{equation}
\left(
\begin{array}{c}
\mbox{\boldmath$\xi$} \\
\tau 
\end{array}
\right) \; = \;
\left(
\begin{array}{c}
\displaystyle
\frac{\mbox{\boldmath$r$}}
{c _{\rm 0} \, \vert t \, - \, t _0 \vert } 
\\
- \, \ln \, \vert 1 \, - \, t / t _0 \vert 
\end{array}
\right) \; ,
\end{equation}
where $ c _0 $ denotes a standard sound speed and
is a function of time $ t $.  The symbol, $ t _0 $, denotes an
epoch at the instant of the protostar formation.
The density in the zooming coordinates,
$ \varrho $, is related with that in the ordinary coordinates,
$ \rho $, by
\begin{equation}
\varrho (\mbox{\boldmath$x$}, \, \tau) \; = \;
4 \pi G \, \rho \, (t \, - \, t _0 ) ^2 \; .
\label{density}
\end{equation}
We define the standard sound speed, $ c _0 $, so that it denotes the
sound speed at a given $ t $ when $ \varrho \; = \; 1 $.  Thus 
it is expressed as
\begin{equation}
c _0 \; = \; \sqrt{\gamma K} \, ( 4 \pi G ) ^{(1 \, - \, \gamma) \,
/ \, 2} \, \vert t \, - \, t _0 \vert ^{1 \, - \, \gamma} \; .
\end{equation}
The pressure in the zooming coordinates, $ p $, is related with
the that in the ordinary coordinates, $ P $, by
\begin{equation}
p \; = \; \frac{4 \pi G }{c _0 {}^2} \,
 P \, (t \, - \, t _0) ^2 \; . \label{pressure}
\end{equation}
Substituting equations (\ref{density}) and (\ref{pressure})
into equation (\ref{polytrope}), we obtain the polytrope
relation in the zooming coordinates,
\begin{equation}
p \; = \; \frac{ \varrho ^\gamma }{\gamma} \; .
\label{polytrope2}
\end{equation}
The velocity in the zooming coordinates, 
$ \mbox{\boldmath$u$} $, is defined as
\begin{equation}
\mbox{\boldmath$u$} \; = \;
\frac{\mbox{\boldmath$v$}}{c _0} \; + \;
(2 \, - \, \gamma) \, 
\frac{\mbox{\boldmath$r$}}{c _0 \, \vert t \, - \, t _0 \vert}
\; .  \label{velocity}
\end{equation}
This velocity denotes that with respect to the zooming coordinates,
and includes the apparent motion, the last term 
in equation (\ref{velocity}).
The gravitational potential in the zooming coordinates,
$ \phi $, is related with that in the ordinary coordinates,
$ \Phi $ by
\begin{equation}
\phi \; = \; \frac{\Phi}{c _0 {}^2} \; .
\end{equation}

In the zooming coordinates, the hydrodyanmical equations are
expressed as
\begin{equation}
\frac{\partial \varrho}{\partial \tau} \; + \;
\nabla ^\prime \cdot (\varrho \mbox{\boldmath$u$}) \; = \;
(4 \, - \, 3 \gamma ) \, \varrho \; , \label{sim1}
\end{equation}
\begin{eqnarray}
\frac{\partial}{\partial \tau} \, 
(\varrho \mbox{\boldmath$u$}) & + &
\nabla ^\prime \cdot
( \varrho \mbox{\boldmath$u$} \otimes \mbox{\boldmath$u$}
) \, + \, \nabla ^\prime p \, + \, \varrho \nabla ^\prime \phi
\nonumber \\
& = & (2 \, - \, \gamma)\, (\gamma \, - \, 1) \, 
\varrho \mbox{\boldmath$\xi$} \, + \,
(7 \, - \, 5 \gamma) \, \varrho \mbox{\boldmath$u$} 
\; , \label{sim2}
\end{eqnarray}
and
\begin{equation}
\Delta ^\prime \, \phi \; = \; \varrho \;  \label{sim3} 
\end{equation}
for $ t \, < \, t _0 $.
The symbols, $ \nabla ^\prime $ and $ \Delta ^\prime $,
denote the gradient and Laplacian in the 
$ \mbox{\boldmath$\xi$} $-space, respectively.

Assuming stationary in the zooming coordinates 
($ \partial / \partial \tau \; = \; 0 $) and the spherical symmetry
($ \partial / \partial \theta \; = \; 
\partial / \partial \varphi \; = \; 0 $), we seek
a similarity solution.  Under these assumptions equations (\ref{sim1}),
(\ref{sim2}), and (\ref{sim3}) reduce to 
\begin{equation}
\frac{\partial u _r}{\partial \xi} \; + \;
\frac{u_r}{\varrho} \, \frac{\partial \varrho}{\partial \xi}
\; = \; (4 \, - \, 3 \gamma) \, - \, \frac{2 u _r}{\xi} \; ,
\label{sim4}
\end{equation}
\begin{eqnarray}
u _r \frac{\partial u _r}{\partial \xi} \; + \;
\frac{1}{\varrho} \, \left( \frac{dp}{d\varrho} \right) \,
\frac{\partial \varrho}{\partial \xi} & + & \hskip -5pt
\frac{\partial \phi}{\partial \xi} \, = \, 
(2 \, - \, \gamma) \, (\gamma \, - \, 1) \, \xi 
\nonumber \\ & + & (3 \, - \, 2 \gamma) \, u _r \; ,
\label{sim5}
\end{eqnarray}
and
\begin{equation}
\frac{\partial \phi}{\partial \xi} \; = \;
\frac{1}{\xi^2} \, \int _0 ^\xi \varrho (\zeta) \, \zeta ^2 \;
d\zeta \; = \; \frac{\varrho u _r}{4 \, - \, 3 \gamma} \; ,
\end{equation}
where $ \xi \, = \, \vert \mbox{\boldmath$\xi$} \vert $.
After some algebra we can rewrite equations (\ref{sim4}) and
(\ref{sim5}) into
\begin{eqnarray}
( \varrho ^{\gamma \, - \, 1} \, - \, u _r {}^2 )
\, \left( \frac{d \varrho}{d \xi} \right) & = & \varrho \,
\left\lbrack - \, \frac{\varrho u _r}{4 \, - \, 3 \gamma} 
\right. \nonumber \\  + \, (2 \, - \, \gamma) (\gamma \, - \, 1) \, \xi 
 & + & \left.
(\gamma \, - \, 1) \, u _r \, + \, \frac{2 u _r {}^2}{\xi} 
\right\rbrack \; , \label{sim6}
\end{eqnarray}
and
\begin{eqnarray}
( \varrho ^{\gamma \, - \, 1} \, - \, u _r {}^2 ) \,
\left( \frac{d u _r}{d\xi} \right) & = &
\frac{ \varrho u _r {}^2 }{4 \, - \, 3 \gamma} \nonumber \\
 - \; (2 \, - \, \gamma) \, (\gamma \, - \, 1) \, \xi u _r & - &
(3 \, - \, 2 \gamma) \, u _r {}^2 \; \nonumber \\
 + \; (4 \, - \, 3 \gamma)
\, \varrho ^{\gamma \, - \, 1} & - & 
\frac{2 u _r}{\xi} \, \varrho ^{\gamma \, - \, 1} \; .
\label{sim7} 
\end{eqnarray}
Equations (\ref{sim6}) and (\ref{sim7}) are singular at the
sonic point, $ u _r {}^2 \, = \, \varrho ^{\gamma \, - \, 1} $.
We obtain the the similarity solution by integrating equations
(\ref{sim6}) and (\ref{sim7}) with the Runge-Kutta method. 
In the numerical integration we used the auxiliary variable of
Whitworth, Summers (1985), $ s $, defined by
\begin{equation}
\frac{\partial \xi}{\partial s} \; = \; 
\varrho ^{\gamma \, - \, 1} \, - \, u _r {}^2 
\; . \label{WSvariable}
\end{equation}
Using equation (\ref{WSvariable}), we rewrite equations (\ref{sim6})
and (\ref{sim7}) into
\begin{eqnarray}
\frac{d \varrho}{d s} & = & \varrho \,
\left\lbrack - \, \frac{\varrho u _r}{4 \, - \, 3 \gamma} \; +
\; (2 \, - \, \gamma) (\gamma \, - \, 1) \, \xi \right. \nonumber \\
& + & \left.
(\gamma \, - \, 1) \, u _r \, + \, \frac{2 u _r {}^2}{\xi} 
\right\rbrack \; , \label{sim8}
\end{eqnarray}
and
\begin{eqnarray}
\frac{d u _r}{d s} & = &
\frac{ \varrho u _r {}^2 }{4 \, - \, 3 \gamma} \; - \;
(2 \, - \, \gamma) \, (\gamma \, - \, 1) \, \xi u _r 
\nonumber \\ & - &
(3 \, - \, 2 \gamma) \, u _r {}^2 \; 
\; + \; (4 \, - \, 3 \gamma)
\, \varrho ^{\gamma \, - \, 1} 
\nonumber \\ & - &
\frac{2 u _r}{\xi} \, \varrho ^{\gamma \, - \, 1} \; ,
\label{sim9} 
\end{eqnarray}
respectively.

Similarity solutions exist for $ \gamma \, < \, 4/3 $.
Figure 1 shows the similarity solutions for $ \gamma $ =
0.9, 1.0, and 1.1.  The solid curves denote $ \varrho $
while the dashed curves denote the infall velocity,
$ - \, u \, + \, (2 \, - \, \gamma ) \, \xi $.
These solutions are the same as those obtained by Yahil (1983)
and Suto, Silk (1988).

\begin{figure}
\epsfxsize=8.0cm
\epsfbox{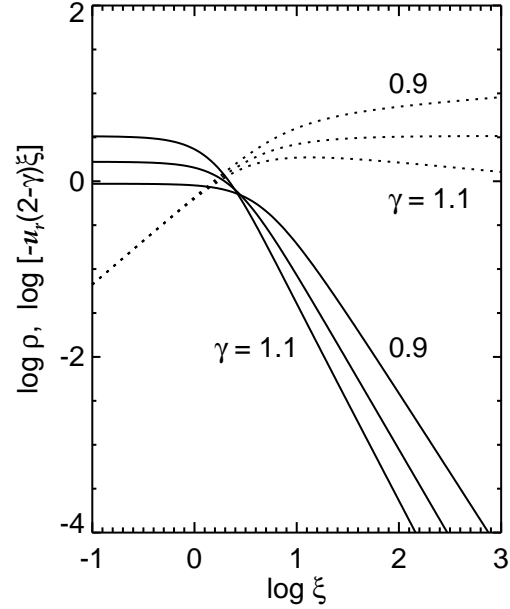}
\caption{The similarity solutions are shown for 
$ \gamma $ = 0.9, 1.0, and 1.1.
The solid curves denote the density, $ \varrho $.
The dashed curves denote the infall velocity,
$ - \, u \, + \, (2 \, - \, \gamma) \, \xi $.
The infall velocity does not include the apparent
motion in the zooming coordinates and is positive
for inward flow.}
\end{figure}

\section{Bar Mode}

In this section we consider a non-spherical perturbation
around the similarity solution.  
The density perturbation is assumed to be proportional
to the spherical harmonics, $ Y _\ell ^m (\theta, \, \varphi) $.
Then the density and velocity are expressed as 
\begin{eqnarray}
\varrho & = &
\varrho _0 \, + \, \delta \varrho (\xi) \, e ^{\sigma \tau}
\, Y _\ell ^m (\theta, \, \varphi) \; , \label{eigen-density} \\
u _r & = & u _{r0} \, + \, \delta u _r (\xi) \, e ^{\sigma \tau}
\, Y _\ell ^m (\theta, \, \varphi) \, , \label{per1} \\
u _{\theta} & = & \delta u _{\theta} (\xi) \, 
\frac{e ^{\sigma \tau}}{\ell \, + \, 1}
\frac{\partial}{\partial \theta} Y _\ell ^m (\theta, \, \varphi)
\, , \label{per2} \\
u _{\varphi} & = & \delta u _{\theta} (\xi) \, 
\frac{e ^{\sigma \tau}}{(\ell \, + \, 1) \, \sin \, \theta}
\frac{\partial}{\partial \varphi} Y _\ell ^m (\theta, \, \varphi)
\, , \label{per3} \\
\phi & = & \phi _0 \, + \, \delta \phi (\xi) \, e ^{\sigma \tau} \,
Y _\ell ^m (\theta, \, \varphi) \; 
\label{per4} \; ,
\end{eqnarray}
where the symbols with suffix, 0, denote the values in the similarity
solution and the symbols with the symbol, $ \delta $, 
denote the perturbations.
Substituting equations (\ref{per1}) throughout (\ref{per4}) 
into equations (\ref{sim1}), (\ref{sim2}), and
(\ref{sim3}), we obtain the perturbation equations,
\begin{eqnarray}
(\sigma \, + \, 3 \gamma \, - \, 4) \, \delta \varrho & + &
\frac{1}{\xi ^2} \, \frac{\partial}{\partial \xi} \,
\lbrack \xi ^2 \, ( \varrho _0 \, \delta u _r \, + \, u _{r0} \delta \varrho) 
\rbrack \nonumber \\
& - & \ell \, \frac{\varrho _0 \delta u _\theta}{r}
\, = \, 0 \; ,
\end{eqnarray}
\begin{eqnarray}
(\sigma \, + \, 2 \gamma \, - \, 3) \, \delta u _r
& + & \frac{\partial}{\partial \xi} \, 
(u _{r0} \delta u _r ) \, + \,
\frac{\partial}{\partial \xi} \, \left( \frac{ \delta \varrho }
{\rho _0 ^{2 \, - \, \gamma} } \right) \nonumber \\
& + & \delta \Gamma \, = \, 0 \; ,
\end{eqnarray}
\begin{eqnarray}
(\sigma \, + \, 2 \gamma \, - \, 3) \, \delta u _\theta & + &
\frac{u_{r0}}{\xi} \, \frac{\partial}{\partial \xi} \, 
(\xi \delta u _\theta) \nonumber \\  + \,
\frac{\ell \, + \, 1}{\xi} \, \left( \frac{\delta \varrho}
{\varrho _0 {}^{2 \, - \, \gamma}} \right. & + & \left. \frac{}{}
\delta \phi
\right) \, = \, 0 \; ,
\end{eqnarray}
\begin{equation}
\frac{\partial}{\partial \xi} \, \delta \phi \, = \, \delta \Gamma \; ,
\label{perg1}
\end{equation}
and
\begin{equation}
\frac{\partial}{\partial \xi} \, \delta \Gamma \, = \,
- \, \frac{2 \, \delta \Gamma}{\xi} \, + \,
\frac{\ell \, ( \ell \, + \, 1)}{\xi ^2} \, \delta \phi \,
+ \, \delta \varrho \; . \label{perg2} 
\end{equation}
These perturbation equations have singularities at the
origin ($ \xi \, = \, 0 $), the sonic point 
$ \lbrack (u _{r0}) ^2 \, - \, \varrho ^{\gamma \, - \, 1} 
\rbrack $, and the infinity ($ \xi \, = \, + \infty $).
These perturbation equations
reduce to those of Hanawa, Matsumoto (1999) when 
$ \gamma \, = \, 1 $.  
We solve these perturbation equations as an eigenvalue problem
for the growth rate, $ \sigma $, according to them.  The
details of the numerical procedures are given in Hanawa,
Matsumoto (1999).  The growth rate is obtained as a function
of $ \gamma $ and $ \ell $. The growth is independent of
$ m $ since the unperturbed state is spherically symmetric
(see, e.g., Hanawa, Matsumoto 1999).

Figure 2 shows the growth rate, $ \sigma $, for 
$ \ell \, = \, 2 $ mode as a function of $ \gamma $.  
The open squares denote the numerically obtained growth rates
and the curve labeled \lq \lq Bar'' denotes a smooth fit to
them.  The growth rate is larger for smaller $ \gamma $.
A collapsing gas sphere is unstable only when 
$ \gamma \, < \, 1.097 $.  We cannot find a damped mode since
appropriate boundary condition are not given at 
$ \xi \, = \, \infty $ for it (see appendix 2).

\begin{figure}
\epsfxsize=8.0cm
\epsfbox{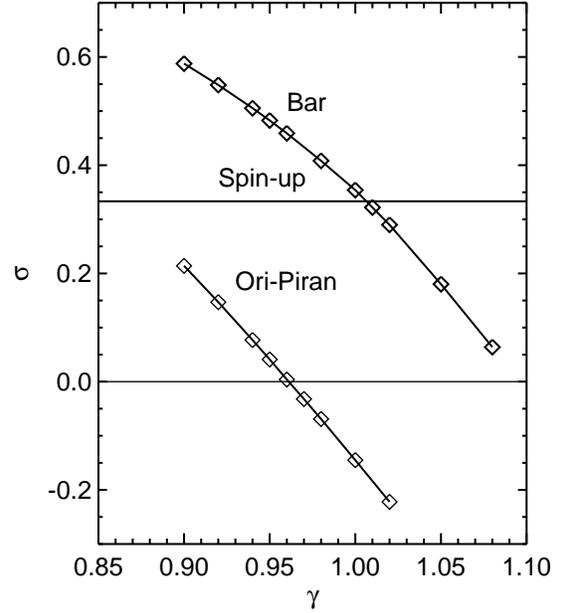}
\caption{The growth rate, $ \sigma $, is shown 
as a function of $ \gamma $.
The open squares denote numerically obtained data points.}
\end{figure}

We searched for $ \ell \, = \, 3 $ mode but could find none.
Although $ \ell \, = \, 1 $ mode exists, it denotes only the
misfit to the center of gravity and is not relevant to
real instability (Hanawa, Matsumoto 1999).  
A higher $ \ell $ mode is also unlikely to exist.  
Thus only the bar mode gives a 
non-spherical density perturbation growing during the
collapse.
Since the bar mode is stabilized for $ \gamma \, > \, 1.097 $,
a similarity solution for $ \gamma \, > \, 1.097 $ is stable
against any non-spherical density perturbation.  This result
can answer to the question posed by Goldreich, Lai, 
Sahrling (1997).  They asked whether a dynamically collapsing
iron resuting type II super nova is unstable against
a large scale non-spherical perturbation.  Since the dynamically
collapsing iron core is well approximated by a similarity
solution for $ \gamma \, \simeq \, 4/3 $, it is stable
against any non-spherical density perturbation.

Figure 3 illustrates the variety of the bar mode.  
Each panel shows a dynamically collapsing isothermal
gas cloud suffering a different bar mode perturbation.
It shows the density distribution perspectively 
by isodensity surfaces. Panel (a) denotes a bar mode
of $ m \, = \, 0 $.  In panel (a) the collapsing gas
cloud is elongated in the $ z $-direction.
Panel (b) denotes the same bar mode of $ m \, = \, 0 $
but having the opposite sign.  In panel (b) the collapsing
gas cloud is oblate and compressed in the $ z $-direction.
Panels (c) and (d) denote the bar modes of $ m $ = 1 and 2,
respectively.  When the modes grow, the collapsing gas
cloud becomes triaxial.

\begin{figure}
\epsfxsize=8.0cm
\epsfbox{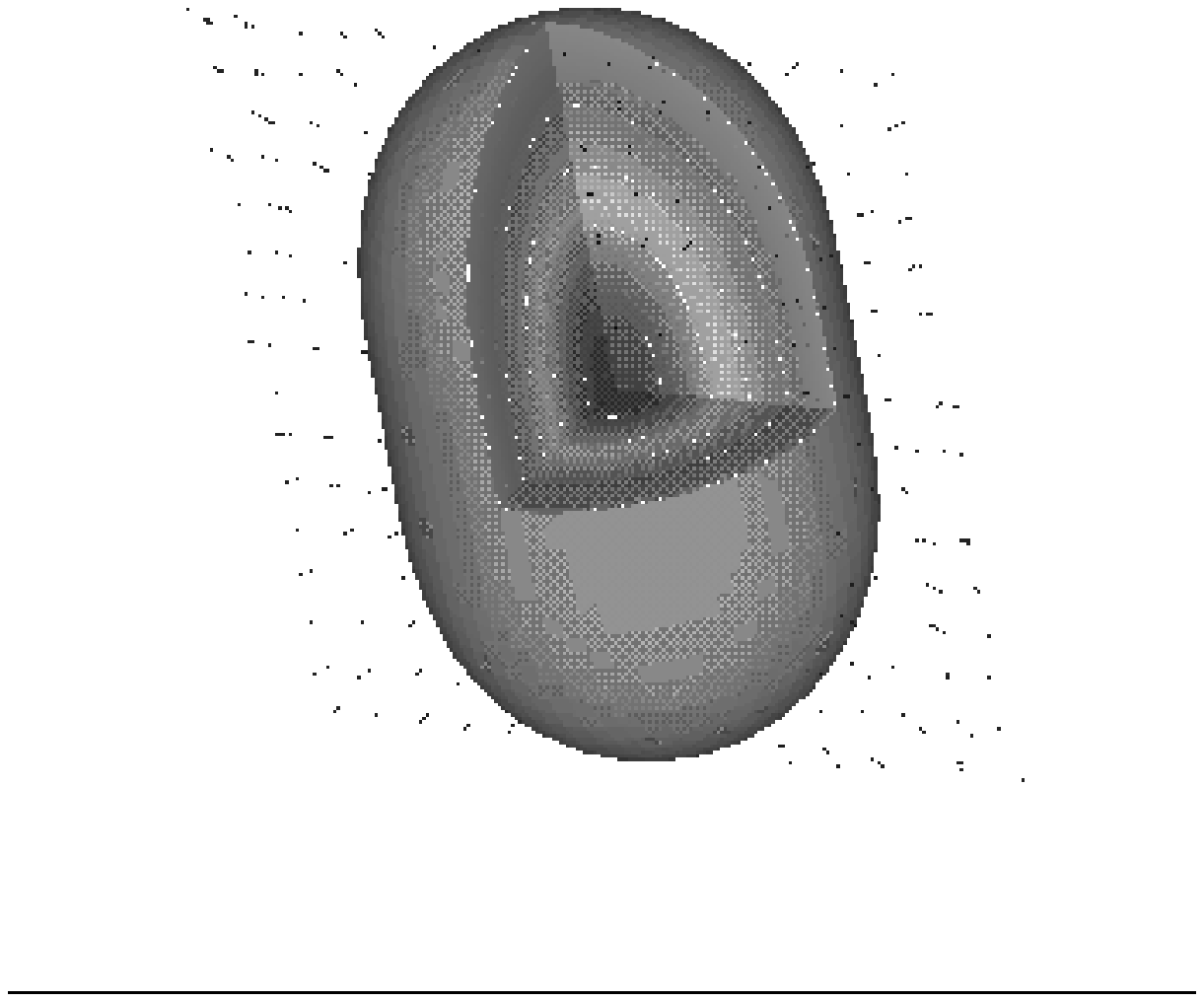}
\epsfxsize=8.0cm
\epsfbox{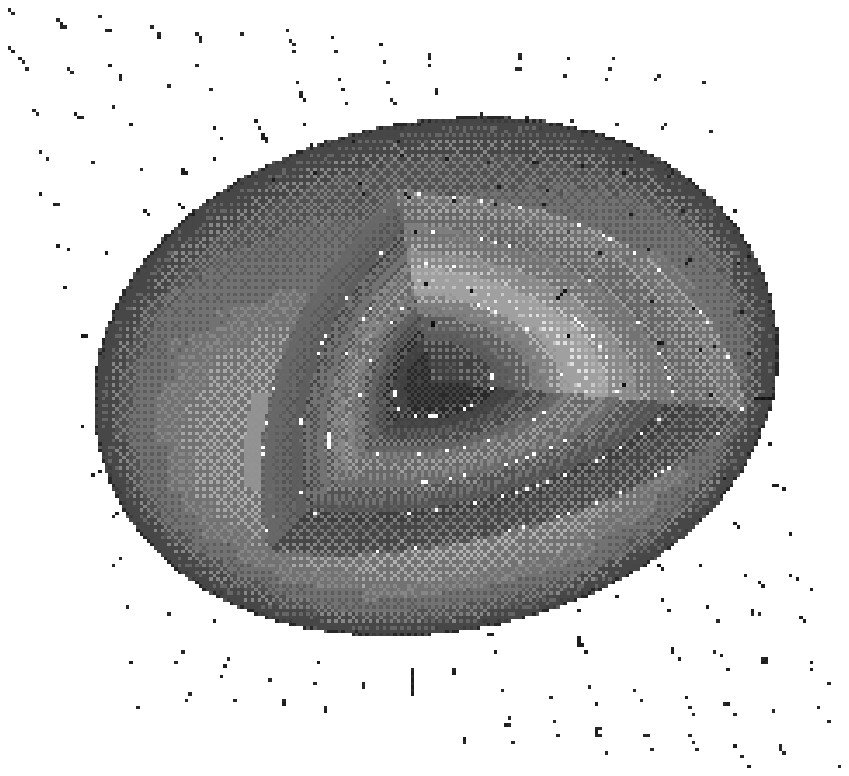}
\caption{A collapsing isothermal cloud core suffering the bar
mode.  The density distribution is shown by the isodensity
surfaces.  The velocity in the $ x - z $ plane is shown
by arrows.  Each panel shows a different bar mode.
Panel (a) denotes the bar mode of $ m \, = \, 0 $.  Panel
(b) denotes the same bar mode of $ m \, = \, 0 $ but having
the opposite sign.  Panel (c) denotes the bar mode of
$ m \, = \, 1 $ while panel (d) does that of $ m \, = \, 2 $.
In panels (a) and (b) the perturbation is added so that
the radial velocity is $ v _{r} \, \propto \,
\lbrack -2/3 \, \pm \, 0.10 \, (3 \cos ^2 \theta \, - \, 1) 
\rbrack \, r $ near the center,
respectiveley.  In  panel (c) the radial velocity is
$ v _r \, \propto \, ( -2/3 \, + \, 0.05 \,
\sin 2 \theta \, \cos \varphi ) \, r $ near the center.
In panel (d) it is 
$ v _r \, \propto \, ( -2/3 \, + \, 0.10 \,
\sin^2 \theta \, \cos 2 \varphi ) \, r $ near the center.}
\end{figure}

\begin{figure}[ht]
\epsfxsize=8.0cm
\epsfbox{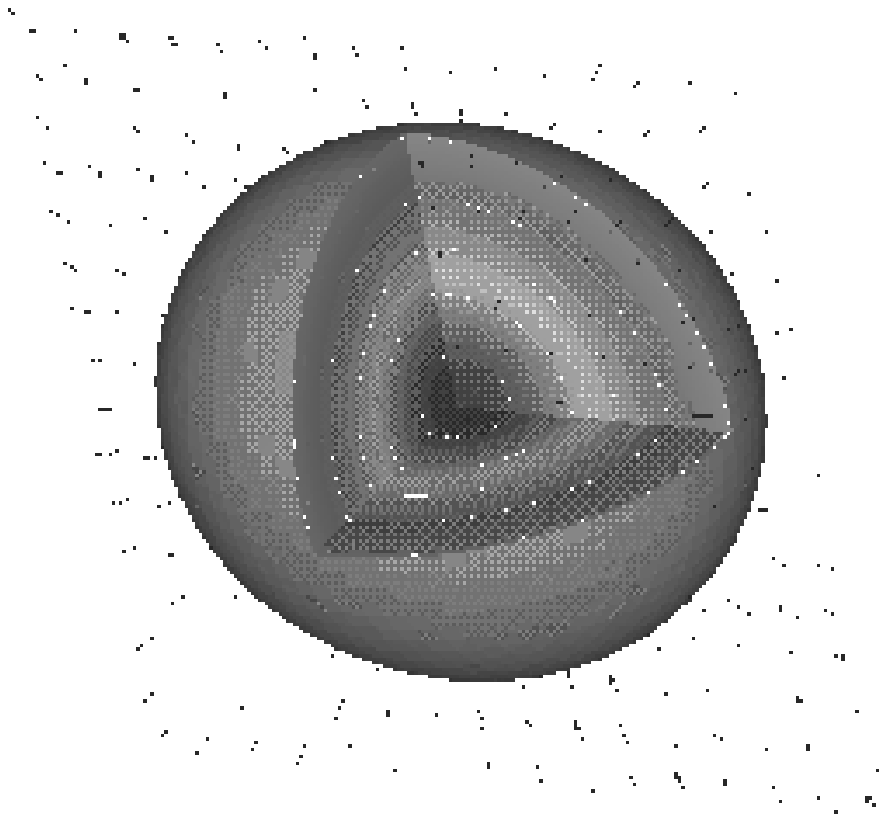}
\epsfxsize=8.0cm
\epsfbox{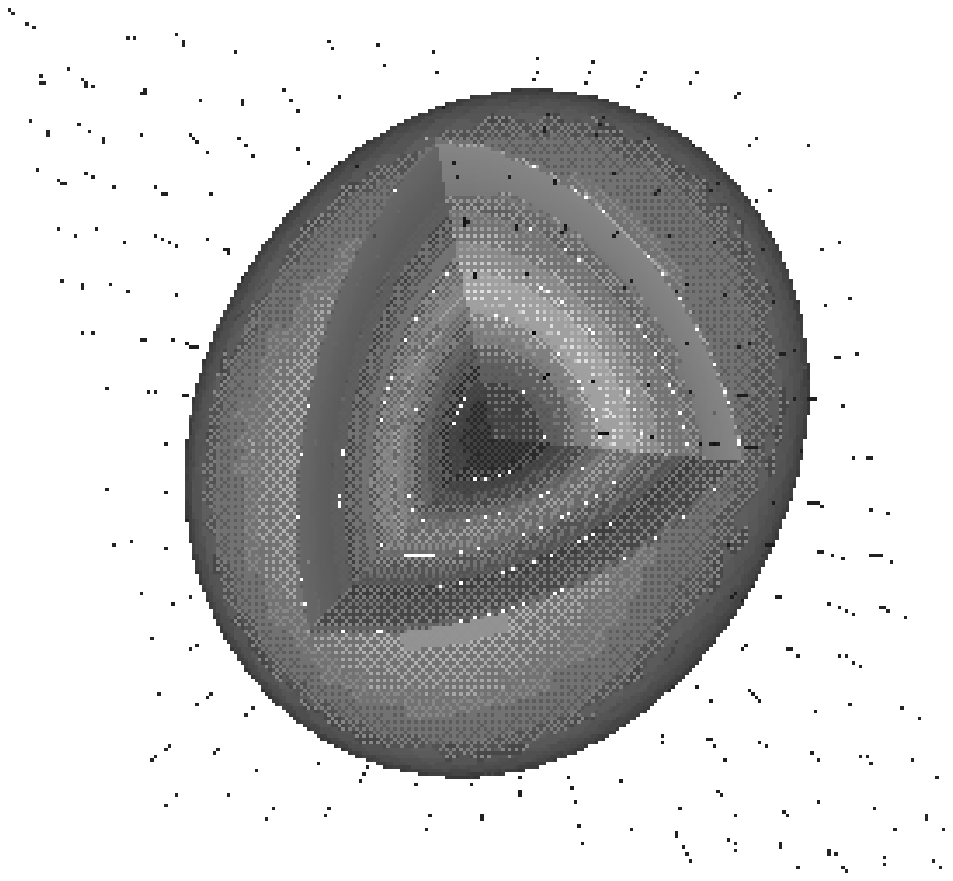}
\end{figure}

\section{Spin-up Mode}

In this section we consider the velocity perturbation expressed as
\begin{equation}
\left(
\begin{array}{c}
u _r \\ u _\theta \\ u _\varphi
\end{array}
\right)
\; = \; 
\left\lbrack
\begin{array}{c}
u _{r0} \\  \displaystyle \frac{A _{\ell, \, m} (\xi)}{\xi \, \sin \theta} \,
\frac{\partial}
{\partial \varphi} \, Y _\ell ^m (\theta, \, \varphi) \,
e ^{\sigma \tau} 
\\ - \, \displaystyle \frac{A _{\ell, \, m} (\xi)}{\xi} \, 
\frac{\partial}{\partial \theta} \,
Y _\ell ^m (\theta, \, \varphi) \, e ^{\sigma \tau} \; 
\label{spin1}
\end{array} 
\right\rbrack
\end{equation}
Substituting equation (\ref{spin1}) into equation (\ref{sim1})
we obtain
\begin{equation}
\delta \varrho \; = \; 0 \; , \label{spin2}
\end{equation}
for this mode.  Similarly we obtain 
\begin{equation}
\delta \phi \; = \; 0 \; , \label{spin3} 
\end{equation}
by substituting equation (\ref{spin2}) into equation
(\ref{sim3}).   Substituting equations (\ref{spin1}),
(\ref{spin2}), and (\ref{spin3}) into equation (\ref{sim2}),
we obtain
\begin{equation}
(\sigma \, - \, 7 \, + \, 5 \gamma) \,
\varrho _0 A _{\ell,m} \; + \, 
\frac{1}{\xi ^2} 
\frac{\partial}
{\partial \xi} (\xi ^ 2 \varrho _0 u _{r0} A_{\ell,m} )
\; = \; 0 \; 
\label{spin4} \; . 
\end{equation}
Substituting equation (\ref{sim4}) into equation (\ref{spin4}),
we obtain
\begin{equation}
(\sigma \, - \, 3 \, + \, 2 \gamma) \, A _{\ell,m} \; + \, 
u _{r0} \,
\frac{\partial}
{\partial \xi} ( A_{\ell,m} )
\; = \; 0 \; 
\label{spin5} \; . 
\end{equation}
Substituting equation (\ref{sol2}) into equation 
(\ref{spin5}) we obtain
\begin{eqnarray}
\sigma & = & 3 \, - \, 2 \gamma \, + \,
\left( \gamma \, - \,  \frac{4}{3} \right)
\, \frac{d \ln A _{\ell,m}}
{d \ln \xi} \bigg\vert _{\xi \, = \, 0}  \; \nonumber \\
& = & \frac{1}{3} \; + \;
\left( \gamma \, - \,  \frac{4}{3} \right)
\, \left( \frac{d \ln A _{\ell,m}}
{d \ln \xi} \bigg\vert _{\xi \, = \, 0} \, - \, 2 \right) 
\; .
\end{eqnarray}
The growth rate, $ \sigma $, should be smaller than 1/3 
since the velocity perturbation is regular at the origin only
for $ d \ln A _{\ell,m} / d \ln \xi \, \ge \, 2 $
[see equation (\ref{spin1})].  When $ d \ln A _{\ell,m} /
d \ln \xi \, = \, 2 $, the angular velocity is nearly constant
around the origin.  The growth rate of the spin-up mode,
$ \sigma $ = 1/3 is also shown in Figure 2 for comparison
with that of the bar mode.
This spin-up mode is essentially the same as the spin-up mode
shown in Appendix of Hanawa, Nakayama (1997).

\section{Ori-Piran Mode}

In this section we consider a perturbation emanating from
the sonic point.  According to Ori, Piran (1988) we 
analyze a spherical perturbation which vanishes inside the
sonic point.  Substituting $ \ell \, = \, 0 $ into 
equations (\ref{per1}) through (\ref{per4}) we obtain 
\begin{equation}
\frac{\partial}{\partial \tau} \delta \varrho \, + \,
\frac{1}{\xi ^2} \, \frac{\partial}{\partial \xi} \,
\lbrack \xi ^2 \, ( \delta \varrho  u _{r0} \,
+ \, \varrho _0 \delta u _r ) \rbrack \, = \,
(4 \, - \, 3 \gamma) \, \delta \varrho \; , 
\label{mass1}
\end{equation}
\begin{eqnarray}
\frac{\partial}{\partial \tau}\delta u _r & + &
\frac{\partial u _{r0}}{\partial \xi} \, \delta u _r \;
+ \; u _{r0} \, \frac{\partial}{\partial \xi} \delta u _r
\nonumber \\& - & (\varrho _0) ^{\gamma \, - \, 3} \,
\frac{\partial \varrho _0}{\partial \xi} \, \delta \varrho 
\, + \, \frac{1}{\varrho _0} \, \frac{\partial}{\partial \xi} \,
\lbrack (\varrho _0) ^{\gamma \, - \,1} \, \delta \varrho \rbrack 
\nonumber \\ &  + &
\delta \Gamma \; = \; (3 \, - \, 2 \gamma) \, \delta u _r \; ,
\label{momentum1}
\end{eqnarray}
and
\begin{equation}
\delta \Gamma \; = \; \frac{1}{\xi^2} \, \int _0 ^\xi \,
\delta \varrho (\zeta) \, \zeta^2 \, d\zeta \; .
\label{gravity1}
\end{equation}
We evaluate equations (\ref{mass1}), (\ref{momentum1}), and
(\ref{gravity1}) at the sonic point.  Then we obtain
\begin{equation}
\varrho _0 \, \frac{\partial}{\partial \xi} \delta u _r
\; + \; u _{r0} \, \frac{\partial}{\partial \xi} \delta \varrho
\; = \; 0 \label{wave1} \; , 
\end{equation}
\begin{equation}
u _{r0} \, \frac{\partial}{\partial \xi} \delta u _r \; + \;
(\varrho _0) ^{\gamma \, - \, 2} \, \frac{\partial}{\partial \xi}
\delta \varrho \; = \; 0 \; , \label{wave2}
\end{equation}
\begin{equation}
\delta \phi \; = \; 0 \; ,
\end{equation}
at the sonic point.  Equations (\ref{wave1}) and (\ref{wave2})
are equivalent and indicate that the sound wave traveling outwards 
with the phase speed $ u _r \, + \, c _{\rm s} $ vanishes.
In other words the perturbation emanating from the sonic point
is the other sound wave of which phase velocity is zero at
the sonic pint, $ \, u _r \, - \, c _{\rm s} \, = \, 0 $.
Taking the linear combination of the $ \xi $-derivatives of
equations (\ref{mass1}) and (\ref{momentum1}) and evaluating
it at the sonic point, we obtain
\begin{eqnarray}
\frac{\partial}{\partial \tau} \,
\left( u _{r0} \, \frac{\partial}{\partial \xi} \delta \varrho
\, - \, \varrho _0 \, \frac{\partial}{\partial \xi} \delta u _r \right)
& - & \left\lbrack - \, 2 \, \frac{\partial u_{r0}}{\partial \xi}
\right. \nonumber \\ 
+ \, (\gamma \, - \, 1) \, u _{r0} \,
\frac{\partial \ln \varrho _0}{\partial \xi} & + & \left.
\left( \frac{ 7 \, - \, 5 \gamma}{2} \right) \right\rbrack \; 
\nonumber \\
\times \, \left( u _{r0} \, \frac{\partial}{\partial \xi} \delta \varrho
\, - \, \varrho _0 \, \frac{\partial}{\partial \xi} \delta u _r \right)
& = & 0 \label{growth2}
\end{eqnarray}
where equations (\ref{wave1}) and (\ref{wave2}) are substituted.
Equation (\ref{growth2}) means
\begin{equation}
\sigma \; = \; - \, 2 \, \frac{\partial u_{r0}}{\partial \xi}
\, + \, (\gamma \, - \, 1) \, u _{r0} \,
\frac{\partial \ln \varrho _0}{\partial \xi} \, + \,
\left( \frac{ 7 \, - \, 5 \gamma}{2} \right) \label{equation}
\; . \label{growth3}
\end{equation}
When $ \gamma \, = \, 1 $, equation (\ref{growth3}) is 
equivalent to equation (15).  To elucidate the growth
(or damping) of this mode we rewrite 
equation (\ref{growth3}) into
\begin{equation}
\sigma \; = \; - 2 \, 
\frac{\partial}{\partial \xi} \,
(  u_{r0} \, - \, c_{\rm s} )
\, + \, \left( \frac{ 7 \, - \, 5 \gamma}{2} \right) 
\; . \label{growth4}
\end{equation}
The first term in the right hand side of equation 
(\ref{growth4}) denotes the dilution of the wave 
due to the spatial variation of the phase velocity.
The phase speed vanishes at the sonic point and increases
with the radial distance.  Thus the wave dilutes and the
amplitude decreases.  The second term in  
equation (\ref{growth4}) denotes the self-reproduction.
As shown in the right hand side of 
equations (\ref{sim1}) and (\ref{sim2}),
the density and momentum density reproduce and amplify 
themselves when measured in the zooming coordinates.
When the self-reproduction overcomes the dilution, 
the collapsing gas sphere is unstable against the
Ori-Piran mode.

Nonlinear growth of the Ori-Piran mode has not been studied yet.
This is the first report that a similarity solution of
collapsing gas sphere can be unstable against the Ori-Piran
mode.  As far as we know, Ogino, Tomisaka, Nakamura (1999)
is the first report on collaps of a polytropic gas sphere
with $ \gamma \, < \, 1 $.  They reported a little on their
numerical simulations of $ \gamma \, < \, 1 $ and mentioned
nothing on the Ori-Piran instability.
Since the similarity solution is unstable, their numerical
solution may not be well approximated by it.

\section{Summary}

As shown in the previous sections a gravitationally
collapsing polytropic gas sphere can suffer from three
types of instability.  When $ \gamma \, < \, 4/3 $,
it is unstable against spin-up mode shown in section 5.
The spin-up mode grows in proportion to 
$ \vert t \, - \, t _0 \vert ^{-1/3} $ and accordingly
in proportion to $ \rho ^{1/6} $.
When $ \gamma \, < \, 1.097 $, it is unstable also
against the bar mode instability.  The growth rate
of the bar mode is large for smaller $ \gamma $.
When $ \gamma \, < \, 1.006 $, it is larger than
that of the spin-up mode.  When $ \gamma \, \le \, 0.961 $,
a gravitationally collapsing polytropic gas sphere is
unstable also against the Ori-Piran mode.

It is particularly interesting that the bar mode has
a larger growth rate when $ \gamma $ is smaller.  
This means that the bar mode grows faster when the
sound speed decreases with the increase in the density.
This result may be valid also for a gravitationally collapsing
gas sphere in which turbulence has an effective pressure.
Since the effective sound speed is lower in a denser part
of molecular cloud, a collapsing cloud core may be
more unstable against the bar mode than a purely isothermal
gas sphere.  This enhancement in the bar mode instability
may be relevant to fragmentation of a collapsing cloud core
and formation of multiple stars during the collapse.

\vspace{1pc}\par
We thank  Naoya Fukuda and Kazuya Saigo for
useful discussion and their help for making figures.
This research is financially supported
in part by the Grant-in-Aid for Scientific Research (C) of
the Ministry of Education, Science, Sports and Culture
(No. 09640318) and by the Grant-in-Aid for
Scientific Research on Priority Areas of 
the Ministry of Education, Science, Sports and Culture of Japan
(No. 10147105).

\section*{Appendix 1. \ Similarity Solution for Collapse of 
Polytropic Gas Sphere}
\setcounter{equation}{0}
\renewcommand{\theequation}{A\arabic{equation}}

In this appendix we summarize the main characteristics of the
similarity solution for collapse of polytropic gas sphere.
See Yahil (1983) and Suto, Silk (1988) for the derivation and
more details. 

When we expand the similarity solution around $ \xi \, = \, 0 $
by the Taylor series,
they are expressed as
\begin{eqnarray}
\varrho _0 (\xi) \; = \; \varrho _0 (0) & - &
\frac{\lbrack \varrho _0 (0) \rbrack ^{2 \, - \, \gamma}}{6 \gamma}
\, \left\lbrack \varrho _0 (0) \, - \, \frac{2}{3} \right\rbrack \,
\xi ^2 \nonumber \\
& + & {\cal O} \, (\xi ^2) \; , \label{sol1}
\end{eqnarray}
and
\begin{eqnarray}
u _r (\xi) & = & \left\lbrack (2 \, - \, \gamma ) 
\, - \, \frac{2}{3} \right\rbrack \, \xi \, + \,
\frac{ \lbrack \varrho _0 (0) \rbrack ^{1 \, - \, \gamma}}{15 \gamma}
\nonumber \\& \times &
\left( \varrho _0 \, - \, \frac{2}{3} \right) 
 \left( \frac{4}{3} \, - \, \gamma \right) \, \xi ^3 \,
+ \, {\cal O} \, ( \xi ^5 ) \; . \label{sol2}
\end{eqnarray}
In the region of $ \xi \, \gg \, 1 $ the similarity
solution has the asymptotic form of
\begin{equation}
\varrho \; \propto \; 
\xi ^{- \, 2 \, / \, (2 \, - \, \gamma) } \; , \label{sol3} 
\end{equation}
and
\begin{equation}
\lbrack u _r \, - \, (2 \, - \, \gamma) \, \xi \rbrack \;
\propto \; \xi ^{(1 \, - \, \gamma) \, / \, (2 \, - \, \gamma)} 
\; . \label{sol4}
\end{equation}

\section*{Appendix 2. \ Asymptotic Behavior of Bar Mode Perturbation}

We can derive the asymptotic forms for the bar mode
perturbation from the requirement that the perturbation is regular
at $ \xi \, = \, 0 $.  The density and velocity perturbations
around $ \xi \, = \, 0 $ are expressed as
\begin{eqnarray}
\delta \varrho & = & \alpha \, \lbrack \varrho _0 (0)  \rbrack
^{2 \, - \, \gamma} \, \xi ^\ell \\
\delta u _r & = & \beta \, \ell \, \xi ^{\ell \, - \, 1} \\
\delta u _\theta & = & \beta \, (\ell \, + \, 1) \,
\xi ^{\ell \, - \, 1} \\
\delta \phi & = & - \, 
\left\{ \, \alpha \, + \, \left\lbrack \sigma \, + \,
2 \gamma \, - \, 3 \right. \right. \nonumber \\
& \; & + \, \left. \left. \frac{\ell \, (4 \, - \, 3 \gamma)}{3}
\right\rbrack \, \beta \right\} \, \xi ^\ell \; , 
\end{eqnarray}
where $ \alpha $ and $ \beta $ are free parameters.  When we
derive the above Taylor series expansion, we use equation
(\ref{sol2}).  The derivation is essentially the same as that shown
in Hanawa, Matsumoto (1999). 

As a boundary condition we assume that the relative density
perturbation, $ \delta \varrho / \varrho _0 $, 
is vanishingly small at infinity, 
$ \xi \, = \, \infty $.  After some algebra we obtain the asymptotic 
relations,
\begin{equation}
\frac{\delta \varrho}{\varrho _0} \; \propto \;
\xi ^{- \sigma /(2 \, - \, \gamma)} 
\end{equation}
\begin{equation}
\delta u _r \; \propto \;
\xi ^{- (\sigma \, + \, \gamma \, - \, 1)/(2 \, - \, \gamma)} \; ,
\end{equation}
\begin{equation}
\delta u _\theta \; \propto \;
\xi ^{- (\sigma \, + \, \gamma \, - \, 1)/(2 \, - \, \gamma)} \; ,
\end{equation}
\begin{equation}
\phi \, \propto \, \xi ^{-(\sigma \, - \, 2 \gamma \, + \, 2)
/(2 \, - \, \gamma)} \; ,
\end{equation}
and
\begin{eqnarray}
\phi & = & \left\lbrack \frac{(\sigma \, - \, 2 \gamma \, + \, 2)
(\sigma \, - \, 3 \gamma \, + \, 4)}{(2 \, - \, \gamma)^2}
\, - \, \ell \, (\ell \, + \, 1) \right\rbrack ^{-1}
\nonumber \\
& \; & r ^2 \, \delta \varrho \; .
\end{eqnarray}
When we derive the above relations, we use equations
(\ref{sol3}) and (\ref{sol4}).  See also Hanawa, Matsumoto  
(1999) for the derivation.

\section*{References}
\small

\re
Bouquet S., Feix M.~R., Fijalkow E., Munier A. 1985,
ApJ 293, 494

\re 
Goldreich P., Lai D., Sahrling, M. 1997,
in Unsolved Problems in Astrophysics, ed. J. Bahcall,
J.~P. Ostriker (Princeton Univ. Press, Princeton) p269

\re
Hanawa, T., Matsumoto, T., 1999, ApJ 521, in press

\re
Mathieu, R.~D. 1994, ARA\& 32, 465

\re
Matsumoto, T., Hanawa, T. 1999, ApJ 521, in press

\re
Ori A., Piran T. 1988, MNRAS, 234, 821

\re 
Ostriker J. 1964, ApJ 140, 1056

\re
Silk J., Suto Y. 1988, ApJ, 335, 295

\re
Suto Y., Silk J. 1988, ApJ 326, 527

\re 
Truelove, J. K., Klein, R. I., McKee, C. F., Holliman, J. H., II,
Howell, L. H., Greenough, J. A. 1997, ApJ 489, L179 

\re
Truelove, J.~K., Klein, R.~I., McKee, C.~F., Hollman, J.~H., II,
Howell, L.~H., Greenough, J.~A., Woods, D.~T. 1998, ApJ 495, 821 

\re
Whitworth A., Summers D. 1985, MNRAS 214, 1

\re 
Yahil A. 1983, ApJ 265, 1047

\label{last}

\clearpage

\end{document}